\documentclass[twocolumn,pra,superscriptaddress,showpacs,10pt,aps]{revtex4-1}
\usepackage[T1]{fontenc}
\usepackage[latin9]{inputenc}
\usepackage{amsmath}
\usepackage{amssymb}
\usepackage{graphicx}
\usepackage{amsthm}
\usepackage{amsfonts}
\usepackage{multirow}
\usepackage{tabularx}
\usepackage{bm}

\begin{document}

\title{Maximal quantum Fisher information for general su(2) parametrization
processes}

\author{Xiao-Xing Jing}
\affiliation{Zhejiang Institute of Modern Physics, Department of Physics, Zhejiang
University, Hangzhou 310027, China}

\author{Jing Liu}
\affiliation{Zhejiang Institute of Modern Physics, Department of Physics, Zhejiang
University, Hangzhou 310027, China}

\author{Heng-Na Xiong}
\affiliation{Department of Applied Physics, Zhejiang University of Technology,
Hangzhou 310023, China}

\author{Xiaoguang Wang}
\email{xgwang1208@zju.edu.cn}
\affiliation{Zhejiang Institute of Modern Physics, Department of Physics, Zhejiang
University, Hangzhou 310027, China}

\begin{abstract}
Quantum Fisher information is a key concept in the field of quantum
metrology, which aims to enhance the accuracy of parameter estimation 
by using quantum resources. In this paper, utilizing a representation
of quantum Fisher information for a general unitary parametrization
process, we study unitary parametrization processes governed by su(2)
dynamics. We obtain the analytical expression for the Hermitian operator
of the parametrization and the maximal quantum Fisher information.
We find that the maximal quantum Fisher information over the parameter
space consists of two parts; one is quadratic in time and the other
oscillates with time. We apply our result to the estimation of a
magnetic field and obtained the maximal quantum Fisher information.
We further discuss a driving field with a time-dependent Hamiltonian
and find that the maximal quantum Fisher information of the driving
frequency attains the optimum when it is in resonance with the atomic frequency.
\end{abstract}

\maketitle

\section{Introduction}

The past two decades have witnessed a dramatic development
of quantum metrology~\cite{Braunstein1994,Giovannetti2004,Giovannetti2006,Roy2008,
Dorner2009,Zwierz2010,MDLang2013,Jarzyna2012,Santarelli1999,Bollinger1996, Joo2011,
Joo2012,BrazJ2011,Escher2011,Demkowicz2012,Chin2012,Alipour2014,Huelga1997},
which is rooted in the theory of quantum parameter estimation. The
estimation of unknown parameters plays an important role in physics and engineering.
By taking advantage of quantum resources such as entanglement and
squeezing, quantum metrology promises a higher precision in parameter
estimation than its classical counterpart. Therefore many practical
applications have been suggested to reap this benefit, including the
detection of gravitational radiation~\cite{McGuirk2002,Braginsky2004,Adhikari2014},
quantum frequency standards~\cite{Huelga1997,Bollinger1996,Santarelli1999}, and
quantum imaging~\cite{Tsang2009,Giovannetti2009,Brida2010}.

Quantum parameter estimation usually involves schemes for optimally estimating properties
of quantum states or processes. The problem of parameter estimation for quantum processes
can be treated as a task of optimal estimation of quantum states, where the states under
investigation are the parameterized output states of the quantum processes. The estimation
of unitary quantum processes is of interest and has been investigated in many different
setups~\cite{Ballester2004,Fujiwara1995}. In a realistic scenario, the system evolution
is unavoidably accompanied by some noise and the quantum process is treated as a
general trace-preserving completely positive map, which is also called a quantum channel.
The optimal estimation of a general one-parameter channel has been studied and some
optimal estimation schemes have been proposed~\cite{Sarovar2006,Fujiwara2003}.  These schemes mainly consist of the identification of an optimal input state and an optimal estimator of the output state. In this paper, we restrict our investigation to unitary quantum processes, for simplicity, and concentrate only on finding the optimal input states and the corresponding accuracy limits.

It is well known that the inverse of Fisher information gives the lower bound of the
accuracy limit. Fisher information characterizes the amount of information about the
true value of a parameter that can be extracted from a probability
distribution.  Quantum Fisher information (QFI), which is a key concept
in quantum metrology, is defined by maximizing the Fisher information
over all possible measurements~\cite{Braunstein1994,Pairs2009}. The quantum
Cramer-Rao theorem asserts that the precision is bounded from below
by the inverse of the QFI~\cite{Helstrom1976,Holevo1982}.
Due to its great importance in quantum metrology and parameter estimation,
the QFI has attracted a lot of attentions~\cite{Petz2002,XMLu2010,Zhang2013,SLLuo2000,Gill2000,
JianMa2011,JingLiuctp2014,XiaoxingJing2014,zhong2013,
Yao2014,ShengshiPang2014,JingLiu2014}.
However, analytical expressions for the QFI are difficult to obtain in
most cases.

The QFI associated with a state~$\rho$~for a parameter~$\theta$
is defined as~\cite{Pairs2009,Helstrom1976,Holevo1982}
\begin{equation}
F=\textrm{Tr}(\rho L^{2}),
\end{equation}
where $\textrm{Tr}$ stands for trace and $L$ is the symmetric logarithmic
derivative operator, which is determined by~$\partial_{\theta}\rho =
\left(L\rho+\rho L\right)/2$.
Consider a general unitary parametrization process~$U=\exp(-itH)$~for an initial pure
state~$|\psi\rangle$, with the time-independent
Hamiltonian~$H$~depending on a parameter~$\theta$, the QFI is simply
connected with the variance of a Hermitian operator~$\mathcal{H}$
in~$|\psi\rangle$, that is~\cite{ShengshiPang2014,JingLiu2014},
\begin{equation}
F=4 \left(\langle\mathcal{H}^2\rangle-\langle \mathcal{H} \rangle^2\right),\label{eq:QFIpure}
\end{equation}
where
\begin{equation}
\mathcal{H}  \equiv  i\left(\partial_{\theta}U^{\dagger}\right)U\label{eq:generator}
\end{equation}
is a Hermitian operator associated with the parametrization.

It has been shown that the variance of an Hermitian operator~$\mathcal{H}$
is maximized when the initial state~$|\psi\rangle$~is an equally
weighted superposition of the eigenvectors~$|\lambda_{M}\rangle$
and~$|\lambda_{m}\rangle$, which correspond to the maximum and minimum
eigenvalues~$\lambda_{M}$~and~$\lambda_{m}$~of the operator~$\mathcal{H}$,
respectively~\cite{Giovannetti2006,ShengshiPang2014}; i.e.,
\begin{equation}
|\psi\rangle=\frac{1}{\sqrt{2}}\left(|\lambda_{M}\rangle+e^{i\phi}|\lambda_{m}\rangle\right), \label{eq:optimalState}
\end{equation}
where~$\phi$~is an arbitrary relative phase, and the maximal QFI (MQFI) is~\cite{ShengshiPang2014}
\begin{equation}
F^{\max}=\left(\lambda_{M}-\lambda_{m}\right)^{2}.\label{eq:max_min}
\end{equation}
Here $|\psi\rangle$ is the optimal input state. The convexity of the QFI guarantees that the MQFI is attained by a pure state rather than mixed states~\cite{Fujiwara2001}. An important observation is that $|\psi\rangle$ can be a maximal entangled state for a many-body system, e.g., GHZ states, or N00N states.
This is consistent with previous research where maximal entangled states are optimal input states to enhance the estimation precision.

Previous research has mainly focused on the estimation of an overall
multiplicative factor of a Hamiltonian~$H$~\cite{Bollinger1996,Jarzyna2012,Joo2011,Joo2012,MDLang2013,
Zhang2013,XiaoxingJing2014}. In such cases, the~$\mathcal{H}$
operator is easily obtained as $\mathcal{H}=-t\partial_{\theta}H$~\cite{JingLiu2014}.
In a recent work \cite{ShengshiPang2014}, Pang \emph{et al.} studied
a general Hamiltonian parameter estimation problem where the parameter
does not appear as an overall multiplicative factor. Their study extends
the quantum metrology to a more general case. Based on their work,
we offered another approach to get the QFI for unitary parametrization
processes in~\cite{JingLiu2014}. In this approach, the QFI
is related to the spectral decomposition of the initial state and
the $\mathcal{H}$ operator of the parametrization process
\begin{equation}
\mathcal{H}=-t\partial_{\theta}H+i\sum_{k=1}^{\infty}\frac{(it)^{k+1}}{(k+1)!}H^{\times k}(\partial_{\theta}H),\label{eq:H}
\end{equation}
where the superoperator $H^{\times k}$ denotes a $k$th-order nested
commutator operation, $H^{\times k}(\cdot)=[H,\cdot\cdot\cdot,[H,\cdot]]$.
When $\theta$ is a multiplicative factor, as in the Mach-Zehnder
interferometer setup, $H=\theta^{k}H',(k=1,2,...)$\cite{Joo2011,Joo2012},
where $H'$ is independent of the parameter to be estimated,
the summation vanishes, and only the first term on the right-hand side
of Eq.~(\ref{eq:H}) contributes.

In this paper, we investigate a general unitary parametrization process
governed by an su(2) Hamiltonian in the form of $\mathbf{r\cdot J}$, where
$\mathbf{r}$ is a time-independent function of the parameter and
$\mathbf{J}$  is the generator of su(2) algebra. This Hamiltonian covers
many interesting applications in physics. We derive the analytical
expression of the MQFI and find that it can be divided into
two parts; one is quadratic in time $t$ due to the dependence
of the norm $|\mathbf{r}|$ on the parameter $\theta$, and the other
oscillates with time $t$ because of the dependence of the direction
$\mathbf{e_r}=\mathbf{r}/|\mathbf{r}|$ on the parameter $\theta$.
Furthermore, we apply the theoretical expression in three practical examples.
The first two examples undergo time-independent su(2)
parametrization, and  the third one is subjected to time-dependent
su(2) parametrization.

This paper is structured as follows. In Sec II, we derive the $\mathcal{H}$
operator and the MQFI for a general su(2) Hamiltonian. In Sec III,
three applications of the theoretical result are listed. The first two are governed by time-independent su(2) Hamiltonians. We discuss the corresponding MQFI for the estimated parameters.
In the third application, we further study the parameter estimation with
a time-dependent su(2) Hamiltonian and give the expression of
the $\mathcal{H}$ operator and the MQFI.
Finally, we give the discussion and conclusion in Sec IV.

\section{Maximal Quantum Fisher information for a general time-independent
su(2) Hamiltonian}

In this section, we investigate the $\mathcal{H}$ operator and the MQFI for
a general su(2) parametrization process. Assume that the Hamiltonian
takes the form of
\begin{equation}
H=\mathbf{r}\cdot\mathbf{J}, \label{eq:Hamiltonian1}
\end{equation}
where $\mathbf{J}=\left(J_{x},J_{y},J_{z}\right)$ is the generator
of su(2) algebra and $\mathbf{r}=\mathbf{r}(\theta)$ is a time-independent
parametric curve in the parameter space. We denote the derivative
of $H$ with respect to $\theta$ as
\begin{equation}
\partial_{\theta}H=\mathbf{v}\cdot\mathbf{J},
\end{equation}
where we define the velocity $\mathbf{v}\equiv d\mathbf{r}/ d\theta$, which denotes the change of $\mathbf{r}$ over the parameter space of $\theta$, including both of the change of $\mathbf{|r|}$ over $\theta$
and the change of the unit vector $\mathbf{e_r}=\mathbf{r/|r|}$ over $\theta$. Thus the velocity
$\mathbf{v}$ can be decomposed as~$\mathbf{v}= \mathbf{v}_{r}+\mathbf{v}_{t}$, with the radial velocity $\mathbf{v}_{r}$ and the transverse velocity~$\mathbf{v}_{t}$~reading as
\begin{subequations} \label{eq:velocities}
\begin{equation}
  \mathbf{v}_{r}  =  \frac{d|\mathbf{r}|}{d\theta}\mathbf{e_r}, \label{eq:radialvelocity}
  \end{equation}
  \begin{equation}
  \mathbf{v}_{t}  =  |\mathbf{r}|\frac{d\mathbf{e_r}}{d\theta}. \label{eq:transverseVelocity}
  \end{equation}
\end{subequations}

Utilizing the commutation relation for su(2) algebra
$
[\mathbf{a}\cdot\mathbf{J},\mathbf{\ b}\cdot\mathbf{J}]=i(\mathbf{a\times b})\cdot\mathbf{J},
$
the first two commutators in Eq.~\eqref{eq:H} read
\begin{subequations} \label{eq:commutator1}
  \begin{align}
    H^{\times}\partial_{\theta}H  &=  i(\mathbf{r\times v})\cdot\mathbf{J}, \\
        \begin{split}
H^{\times2}\partial_{\theta}H & =  -\left[\mathbf{r}\times\left(\mathbf{r}\times\mathbf{v}\right)\right]\cdot\mathbf{J} \\
 & =  [|\mathbf{r}|^{2}\mathbf{v}-(\mathbf{r\cdot v})\mathbf{r}]\cdot\mathbf{J},
        \end{split}
  \end{align}
\end{subequations}
where the relation
$\mathbf{a\times(b\times c)}=\mathbf{(a\cdot c)b-(a\cdot b)c}$ is
used in the last equality. It is shown that both $H^{\times}\partial_{\theta}H$
and $H^{\times2}\partial_{\theta}H$ are eigen operators of the superoperator
$H^{\times 2k}$, i.e.,
\begin{subequations} \label{eq:commutator2}
  \begin{equation}
    H^{\times(2k+1)}\partial_{\theta}H  =  |\mathbf{r}|^{2k}H^{\times}\partial_{\theta}H,
  \end{equation}
  \begin{equation}
    H^{\times(2k+2)}\partial_{\theta}H  =  |\mathbf{r}|^{2k}H^{\times2}\partial_{\theta}H,
  \end{equation}
\end{subequations}
where $k = 1,2,...$.
By plugging Eq.~\eqref{eq:commutator2} 
into Eq.~\eqref{eq:H}, we have
\begin{eqnarray}
\mathcal{H} & = & -t\partial_{\theta}H+i\sum_{k=0}^{\infty}\frac{(it)^{2k+2}}{(2k+2)!}
 H^{\times(2k+1)}(\partial_{\theta}H)                   \nonumber \\
& &+i\sum_{k=0}^{\infty}\frac{(it)^{2k+3}}{(2k+3)!}
 H^{\times(2k+2)}(\partial_{\theta}H)                   \nonumber \\
& = & -t\partial_{\theta}H+i\left(\cos\left(|\mathbf{r}|t\right)-1\right)
 \frac{H^{\times}\partial_{\theta}H}{|\mathbf{r}|^{2}}    \nonumber \\
&  & -\left(\sin\left(|\mathbf{r}|t\right)-|\mathbf{r}|t\right)
 \frac{H^{\times2}\partial_{\theta}H}{|\mathbf{r}|^{3}},     \label{eq:HH}
\end{eqnarray}
and inserting the first two commutators in Eq.~\eqref{eq:commutator1} into Eq.~\eqref{eq:HH}, we find that
\begin{eqnarray}
  \mathcal{H} & = & \frac{\left(\mathbf{r\cdot v}\right)\left(\sin\left(|\mathbf{r}|t\right)
 -|\mathbf{r}|t\right)}{|\mathbf{r}|^{3}}\mathbf{r}\cdot\mathbf{J}
 -\frac{\sin\left(|\mathbf{r}|t\right)}{|\mathbf{r}|}\mathbf{v\cdot J}  \nonumber \\
&  & +\frac{1-\cos\left(|\mathbf{r}|t\right)}{|\mathbf{r}|^{2}}(\mathbf{r\times v})\cdot\mathbf{J}.
\end{eqnarray}
According to Eq.~\eqref{eq:velocities}, we have that $\mathbf{(r\cdot v)r}=|\mathbf{r}|^2 \mathbf{v}_{r}$, $\mathbf{r\times v}=
|\mathbf{r}|\mathbf{e_{r}}\times \mathbf{v}_{t} = |\mathbf{r}|(\mathbf{v}_{r}\times\mathbf{v}_{t})/(d\mathbf{|r|}/d\theta)$;
therefore, $\mathcal{H}$ can be further written as
\begin{eqnarray}
  \mathcal{H} & = & \mathbf{A\cdot J}   \nonumber \\
 & = & \frac{1-\cos(|\mathbf{r}|t)}{|\mathbf{r}|\frac{d |\mathbf{r}|}{d\theta}}(\mathbf{v}_{r}\times\mathbf{v}_{t})\cdot \mathbf{J} -t\mathbf{v}_{r}\cdot \mathbf{J}-\frac{\sin(|\mathbf{r}|t)}{|\mathbf{r}|}\mathbf{v}_{t}\cdot \mathbf{J}
\nonumber \\
   \label{eq:su2_H}
\end{eqnarray}
where
\begin{eqnarray}
\mathbf{A}  =  \frac{1-\cos(|\mathbf{r}|t)}
{|\mathbf{r}|\frac{d|\mathbf{r}|}{d\theta}}
(\mathbf{v}_{r}\times\mathbf{v}_{t}) -t\mathbf{v}_{r}-
\frac{\sin\left(|\mathbf{r}|t\right)}
{|\mathbf{r}|}\mathbf{v}_{t}. \label{eq:A}
\end{eqnarray}
The three terms in Eq.~\eqref{eq:A} are perpendicular to
each other and the norm of $\mathbf{A}$ is
$
|\mathbf{A}|  =  \sqrt{\mathbf{v}_{r}^{2}t^{2}+4 \frac{\mathbf{v}_{t}^{2}}{\mathbf{r}^{2}} \sin^2\left(\frac{|\mathbf{r}|}{2}t \right)
},
$
therefore according to Eq.~\eqref{eq:max_min}, the MQFI over the
parameter $\theta$ is
\begin{eqnarray}
F^{\max} & = & \left[j|\mathbf{A}|-(-j)|\mathbf{A}|\right]^{2}\nonumber \\
 & = & 4j^{2}\left[\mathbf{v}_{r}^{2}t^{2}+ 4\frac{\mathbf{v}_{t}^{2}}{\mathbf{r}^{2}} \sin^2\left(\frac{|\mathbf{r}|}{2}t \right) \right],
\label{eq:maxFisher1}
\end{eqnarray}
where $j$ is the maximal eigenvalue of $J_z$. Equation (\ref{eq:maxFisher1})
shows that the MQFI can be divided into two parts. The first
part is quadratic in time $t$ and is proportional to the square
of the radial velocity $\mathbf{v}_{r}$. The second part oscillates
with time~$t$~and is proportional to the ratio of the square of the
transverse velocity $\mathbf{v}_{t}$ to the square of ${\mathbf{r}}$.
According to Eq.~\eqref{eq:transverseVelocity}, this ratio is equal to the square of
the derivative $d\mathbf{e_r}/d\theta$. Thus the quadratic term is due to
the dependence of the norm $|\mathbf{r}|$ on the parameter $\theta$, while the
oscillation term is due to the dependence of the direction $\mathbf{e_r}$ on
$\theta$.

When~$t$~is small, we can expand Eq.~\eqref{eq:maxFisher1}
in~$t$~to the second order, and the MQFI is simplified to
\begin{equation}
F^{\max}=4j^{2}t^{2}\left[\mathbf{v}_{r}^2+\mathbf{v}_{t}^2\right]=4j^{2}t^{2}\mathbf{v}^{2}.
\end{equation}
That is, the MQFI grows quadratically with time $t$ and
depends only on the norm of $\mathbf{v}$.

If the change of $\mathbf{|r|}$ over the parameter space is 0, i.e.,
$\mathbf{v}_{r}= \frac{d |\mathbf{r}|}{d\theta}\mathbf{e_r}=0$,
then the MQFI is reduced to
\begin{equation}
F^{\max}=16j^{2}\frac{\mathbf{v}^{2}}{\mathbf{r}^{2}}\sin^2\left(\frac{|\mathbf{r}|}{2}t \right).\label{eq:maxFisher2}
\end{equation}
In this case, the MQFI always oscillates with the time so that its value is bounded.
At the time points
 $t=(2k+1)\pi/|\mathbf{r}|,k
= 0,1,2,...$, the MQFI reaches the optimal value,
\begin{equation}
F_{\textrm{op}}^{\max}=16j^{2}\frac{\mathbf{v}^{2}}{\mathbf{r}^{2}}.
\end{equation}

\section{Applications}
In this section,  we apply the theoretical expression in three
practical examples. The dynamics of the first two examples are governed by time-independent su(2)
Hamiltonians, and the third one is governed by a  time-dependent su(2) Hamiltonian.
Also note that the Hamiltonians may depend on several
parameters, and we only consider the single parameter estimation in this paper, i.e., we suppose
that other parameters are known when we estimate a specific parameter.

\subsection{Time-independent Hamiltonians} \label{sec3a}

\emph{Case 1}. Let us consider the Hamiltonian, \eqref{eq:Hamiltonian1}, with
$\mathbf{r}$ in the explicit form in the spherical coordinates,
\begin{equation}
  \mathbf{r}=r\left(\sin\theta\cos\varphi,\sin\theta\sin\varphi,\cos\theta\right),
\end{equation}
where $r$ represents the amplitude of the external magnetic field and
$\theta$, $\varphi$ denotes the direction of the field.
Suppose $\theta$ is the parameter we want to estimate. The velocity vector is
readily obtained as $\mathbf{v} = r\left(\cos\theta\cos\varphi,\cos\theta\sin\varphi,-\sin\theta\right)$.
It is obvious that $\mathbf{v_{r}}=0$. According to Eq.~\eqref{eq:maxFisher2},
the MQFI over $\theta$ is
\begin{equation}
F_{\theta}^{\max}=16j^{2}\sin^2\left(\frac{rt}{2}\right),
\end{equation}
and when $t=(2k+1)\pi/r,k=0,1,2,...$, $F_{\theta}^{\max}$ reaches its optimal
value of $16j^{2}$.

Similarly, if $\varphi$ is the parameter to be estimated, the
MQFI can be derived as
\begin{equation}
F_{\varphi}^{\max}=16j^{2}\sin^{2}\theta \sin^2\left(\frac{rt}{2}\right),
\end{equation}
and when $t=(2k+1)\pi/r,k=0,1,2,...$, and $\theta=\frac{\pi}{2}$, $F_{\varphi}^{\max}$
reaches its optimal value of $16j^{2}$. In both cases, the
MQFI is bounded.

Finally, if $r$ is the parameter to be estimated, then $\mathbf{v}^2 = \mathbf{v}_{r}^2 = 1$ and the second
term in Eq.~\eqref{eq:maxFisher1} vanishes. Thus the MQFI reads
\begin{equation}
  F_{r}^{\max}=4j^{2}t^{2}.
\end{equation}
In this case, the MQFI is unbounded and grows quadratically
with time $t$.

\vspace{1cm}

\emph{Case 2}.
Let us consider an ensemble of $N$ two-level atoms interacting with
a static magnetic field. The Hamiltonian of this system can be written
as
\begin{equation}
H=\omega_{0}J_{z}+\lambda J_{x}, \label{eq:Hamiltonian2}
\end{equation}
where~$\omega_{0}$~is the atomic transition frequency and~$\lambda$~is the Rabi
frequency, which is proportional to the amplitude of the static magnetic field.

Here we find the MQFI of~$\omega_{0}$~first.
Comparing with the previous section, the coefficient vector is $\mathbf{r}=(\lambda,0,\omega_{0})^{T}.$
Denote the norm of~$|\mathbf{r}|$~as~$K=\sqrt{\lambda^{2}+\omega_{0}^{2}}$; then the velocity is
~$\mathbf{v} = d\mathbf{r}/d\omega_{0} = (0,0,1)^T$, the radial velocity~$\mathbf{v}_{r}^2 =
d|\mathbf{r}|/d\omega_{0}^2 = \omega_{0}^2/K^2$, and the transverse velocity~$\mathbf{v}_{t}^2=\mathbf{v}^2-
\mathbf{v}_{r}^2 = \lambda^2/K^2$,
thus the corresponding MQFI is readily obtained as
\begin{eqnarray}
F_{\omega_0}^{\max}  =  4j^{2}\left[\frac{\omega_0^{2}}{K^{2}}t^{2}+\frac{4\lambda^{2}}{K^{4}} \sin^2\left(\frac{Kt}{2}\right)\right].\label{eq:Fomega0}
\end{eqnarray}

\begin{figure}
\centering\includegraphics[width=9.0cm,clip]{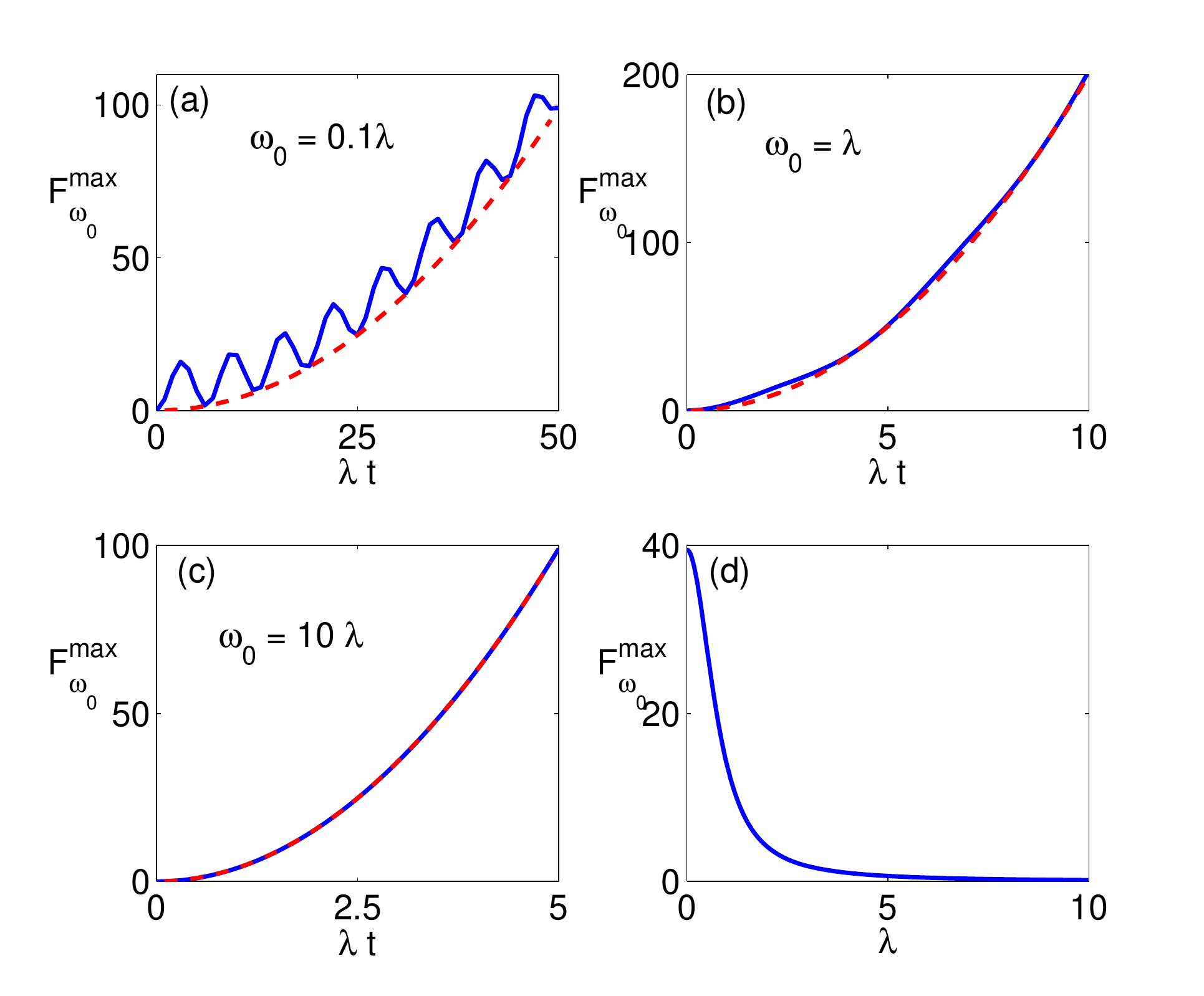}
\caption{\label{figure1}(Color online) The MQFI $F_{\omega_{0}}^{\max}$ as a function of time ~$t$~ with~$\lambda = 1$~ and (a)~$\omega_0$~ = 0.1,(b)~$\omega_0$~ = 1,(c)~$\omega_0$ = 10.
Dashed (red) lines denote the contribution of the first term in Eq.~\eqref{eq:Fomega0} and   solid (blue) lines denote the contribution of both terms. (d) We set~$\omega_0 = 1$~ and~$t = \pi/K$~, and $F_{\omega_{0}}^{\max}$ decreases as $\lambda$ grows. Here we take ~$j = 1$.}
\end{figure}

When $t$  is very large, the oscillating term can be neglected. However,
this term may also be important in some cases. We plot the MQFI $F_{\omega_{0}}^{\max}$ with and without the oscillating term in Fig.~\ref{figure1}. In Fig.~\ref{figure1}(a), where we take $\lambda = 10\omega_{0} = 1$, we see that
the effect of the oscillating term is very strong.
In Fig.~\ref{figure1}(b), we take $\lambda = \omega_{0} = 1$, and the amplitude of
the oscillating term is extremely small in comparison with the quadratic term.
And in Fig.~\ref{figure1}(c), where we take $\lambda = 0.1\omega_{0} = 1$, the
oscillating term could be neglected even when $t$ is very small.

On the other hand, in Fig.~\ref{figure1}(d), we
set $\omega_{0} = 1$ and $t=\pi/K$, and we can see that $F_{\omega_{0}}^{\max}$
decreases with increasing $\lambda$. This can be explained as
follows. When $\lambda$ is very large, the contribution of the first
term $\omega_{0}J_{z}$ in Eq.~\eqref{eq:Hamiltonian2} can be neglected; i.e., when $\lambda\gg\omega_{0}$,
$F_{\omega_{0}}^{\max}\rightarrow 0.$

Following a similar procedure, the MQFI with respect to $\lambda$  is obtained as
\begin{eqnarray}
F_{\lambda}^{\max}  =  4j^{2}\left[\frac{\lambda^{2}}{K^{2}}t^{2}+\frac{4\omega_{0}^{2}}{K^{4}} \sin^2\left(\frac{Kt}{2}\right)\right].\label{eq:Flambda}
\end{eqnarray}
This result is similar to Eq.~\eqref{eq:Fomega0} and we omit the discussion here.

\subsection{Time-dependent Hamiltonian}

Hitherto all the Hamiltonians considered in this paper are time independent.
If the Hamiltonian is time dependent, generally the unitary operator
will involve a time-ordering procedure and Eq.~\eqref{eq:H} fails.
In this subsection, we go beyond the previous situations and investigate
a time-dependent Hamiltonian. In the following we replace the static
magnetic field in the preceding subsection with a time-dependent driving field,
and the Hamiltonian is
\begin{equation}
H = \omega_{0}J_{z}+\lambda\left[J_{x}\cos(\omega t)+J_{y}\sin(\omega t)\right]. \label{eq:Hamiltonian3}
\end{equation}
Here $\omega$ is the frequency of the driving field. This Hamiltonian
could reflect many realistic physical systems,
such as in the Ramsey spectroscopy and NMR techniques~\cite{Ramsey1990,Bollinger1996}.
The~$\mathcal{H}$~operator of this Hamiltonian cannot be written in
the form of Eq.~\eqref{eq:H} and the previous method must be changed.
We can circumvent this problem by utilizing a rotating transform.
In a rotating frame, the original state vector~$|\psi\rangle$~is
transformed as
\begin{equation}
|\tilde{\psi}\rangle=R|\psi\rangle=e^{i\omega tJ_{z}}|\psi\rangle,
\end{equation}
and the effective Hamiltonian in this new frame is
\begin{eqnarray}
H_{\textrm{eff}} & = & RHR^{\dagger}-iR\frac{\partial R^{\dagger}}{\partial t}\nonumber \\
 & = & \Delta J_{z}+\lambda J_{x},
\end{eqnarray}
which is time independent. Here~$\Delta=\omega_{0}-\omega$~denotes
the detuning between the atom transition and the driving magnetic field.

Therefore, the unitary evolution in the original frame is readily
obtained as
\begin{equation}
U=U_1 U_2 = e^{-i\omega tJ_{z}}e^{-iH_{\textrm{eff}}t},\label{eq:unitary2}
\end{equation}
with $U_1 = R^\dagger = e^{-i\omega tJ_z}$ and $U_2 = e^{-iH_{\textrm{eff}} t}$.
In such a way, we turn the time-dependent Hamiltonian of Eq.~\eqref{eq:Hamiltonian3} into two time-independent ones, $H_1 = \omega J_z$, and $H_{\textrm{eff}}$. The $\mathcal{H}$ operator for two consecutive unitary operation $U = U_1 U_2$ can be derived as
\begin{eqnarray}
 \mathcal{H} & = & i\partial_{\theta} \left(U_2^\dagger U_1^\dagger\right)U_1 U_2   \nonumber \\
  & = & \mathcal{H}_2 + U_2^\dagger \mathcal{H}_1 U_2,  \label{eq:H1H2}
\end{eqnarray}
with~$\mathcal{H}_1 = i\partial_{\theta} U_1^\dagger U_1$~and~$\mathcal{H}_2 = i\partial_{\theta} U_2^\dagger U_2$. Here $\theta$ can be $\lambda, \omega_0,$ or $\omega$.
For the case in Eq.~\eqref{eq:unitary2}, if $\lambda$ or $\omega_0$ is the parameter
to be estimated, it is obvious that~$\mathcal{H}_1 = 0$~and one can simply get the
MQFI by replacing~$\omega_0$~in Eq.~\eqref{eq:Fomega0} and ~\eqref{eq:Flambda} with~$\Delta$~. When $\Delta = 0$,
$F_{\lambda}^{\max}$ attains its optimal value of $4j^2 t^2$.

Next, let us find the $\mathcal{H}$ operator with respect to $\omega$.
The Hamiltonian associated with~$\mathcal{H}_2$~is~$H_2 = H_{\textrm{eff}}$,
and the corresponding coefficient vector reads~$\mathbf{r} = (\lambda,0,\Delta)^T$.
Then by utilizing Eq.~\eqref{eq:velocities}, we can obtain that
\begin{eqnarray}
  \mathbf{v}_{r} &=& \frac{\Delta}{K'^2}\left(-\lambda,0,-\Delta\right), \nonumber \\
  \mathbf{v}_{t} & = & \frac{\lambda}{K'^2}\left(\Delta,0,-\lambda \right), \nonumber \\
  \mathbf{v}_{r}\times\mathbf{v}_{t} & = & \frac{-\lambda\Delta}{K'^2}\left(0,1,0\right).
\end{eqnarray}
Therefore, according to Eq.~\eqref{eq:su2_H}, we have that
\begin{eqnarray}
  \mathcal{H}_2 & = & \frac{1}{K'^3}\big[\lambda\Delta \left[K't-\sin(K't)\right]J_x \nonumber\\
  & & +\lambda K'\left[1-\cos(K't)\right] J_y   \nonumber \\
  & &+ \left[\Delta^2 K't + \lambda^2\sin(K't)\right]J_z\big].
  \label{eq:H2}
\end{eqnarray}

On the other hand, the Hamiltonian associated with
$\mathcal{H}_1$ reads $H_1 = \omega J_z$, where the parameter enters
in the Hamiltonian as a multiplicative factor.
According to Eq.~\eqref{eq:H}, we have $\mathcal{H}_1 = -tJ_z$. Based
on the formula $\exp(A)B\exp(-A)=\exp(A^\times)B=\sum_{k=0}^\infty \frac{A^{\times n}}{n!}B$,
we obtain
\begin{eqnarray}
  &   & U_2^\dagger \mathcal{H}_1 U_2 \nonumber \\
  & = & -\frac{t}{K'^2}\big[\lambda\Delta\left[1-\cos(K't)\right] J_x + \lambda K'\sin(K't) J_y  \nonumber \\
  & & +\left[\Delta^2+\lambda^2\cos(K't)\right]J_z \big],  \label{eq:UH1U}
\end{eqnarray}
where we denote~$K' = \sqrt{\lambda^2 + \Delta^2}$.

Combining Eqs.~\eqref{eq:H1H2}, ~\eqref{eq:H2}, and ~\eqref{eq:UH1U},
we have
\begin{eqnarray}
\mathcal{H} & = &\mathbf{A'}\cdot\mathbf{J} \nonumber\\
& = & \frac{\left[\sin\left(K't\right)-K't\cos\left(K't\right)\right]}{K'^{3}}\left(-\lambda\Delta J_{x}+\lambda^{2}J_{z}\right)\nonumber \\
 &  & +\frac{\left[1-\cos\left(K't\right)-K't\sin\left(K't\right)\right]}{K'^{2}}\lambda J_{y}.
\end{eqnarray}
Based on the norm of the vector $\mathbf{A'}$, the MQFI is
readily obtained as
\begin{eqnarray}
F_{\omega}^{\max} &= & 4j^{2}\frac{\lambda^{2}}{K'^{4}}\left[2+K'^{2}t^{2} \right. \nonumber \\
& & \left.-2K't\sin\left(K't\right)-2\cos\left(K't\right)\right].\label{eq:maxFisher3}
\end{eqnarray}

\begin{figure}[tbp]
\centering\includegraphics[width=9.0cm,clip]{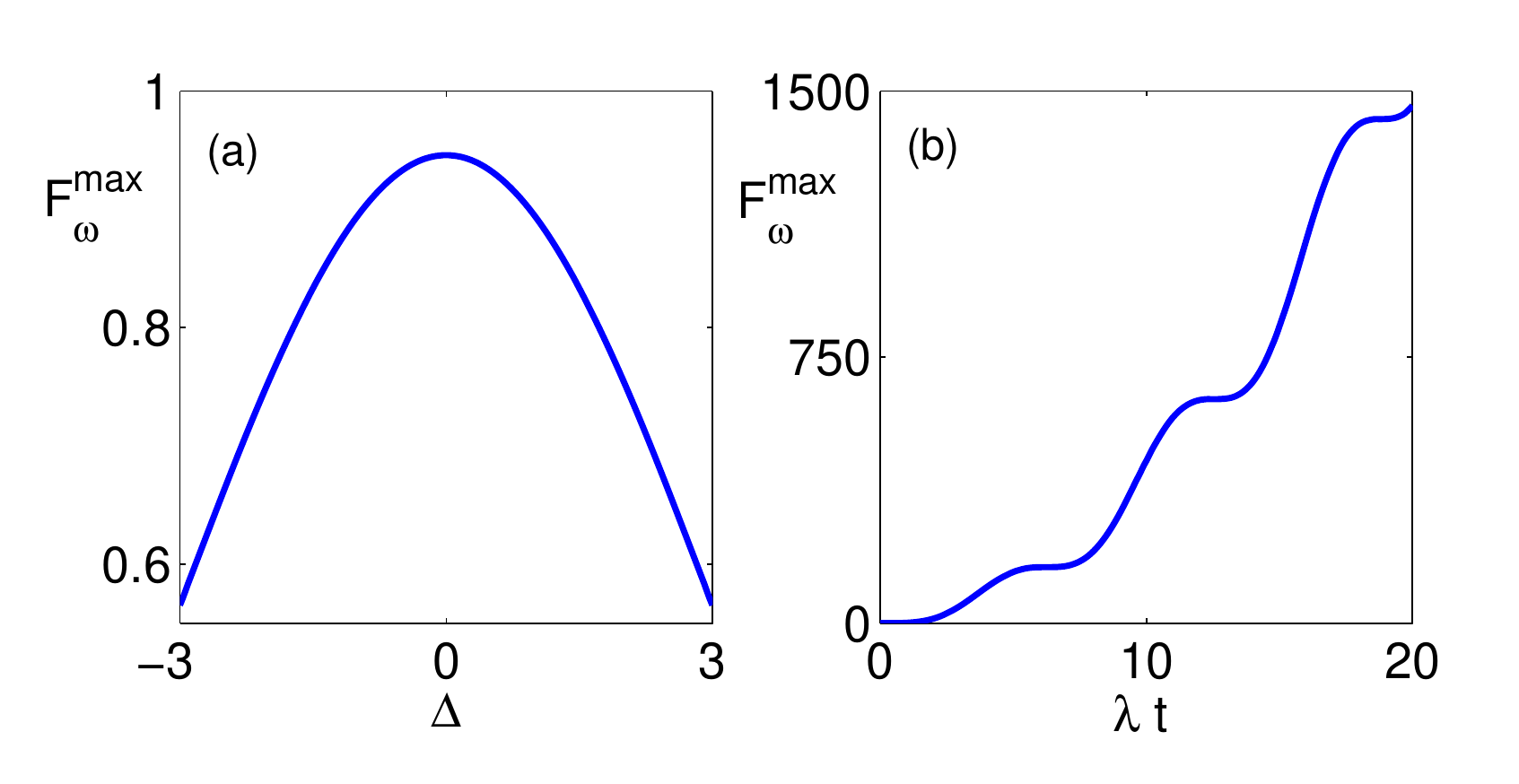}
\caption{\label{figure2} The MQFI $F_{\omega}^{\max}$  as a function of the detuning $\Delta$  (a) and $\lambda$ (b). (a) We take $\lambda = 1$, and $t = 1$. $F_{\omega}^{\max}$ attains the optimum when $\Delta = 0$. (b), We take $\Delta = 0$ and $\lambda = 1$. $F_{\omega}^{\max}$ grows quadratically with time t. We set $j = 1$ in both panels.}
\end{figure}

Figure 2.(a) shows that
when $\Delta=0$ (resonance), $F_{\omega}^{\max}$ attains its optimal
value. This can be proved theoretically by taking the derivative of
Eq. (\ref{eq:maxFisher3}),
\begin{equation}
\frac{d}{d\Delta}F_{\omega}^{\max}|_{\Delta=0}=0.
\end{equation}
When $\Delta = 0$, we plot $F_{\omega}^{\max}$ in Fig.~2(b) with
$\lambda = 1$. As $t$ increases, the MQFI can be approximated
as $F_{\omega}^{\max} = 4j^2 t^2$ and the oscillation can also
be neglected.

\section{Discussion and Conclusion}

In the previous sections, we have only discussed the single-parameter estimation problem. The extension to multi-parameter estimation is straightforward and has been discussed in our recent work \cite{JingLiu2014}.

The quantum Cram\'er-Rao theorem asserts that the variance of any unbiased estimator is bounded
below by the inverse of the QFI. In this paper, we compute the
MQFI for su(2) parametrization processes, thus getting the
corresponding ultimate estimation bound. The achievement of the bound is of practical interest
and various methods of building the optimal estimators are discussed in the literature,
including Bayes estimators and maximum-likelihood estimators \cite{Teklu2009,Brivio2010}. The optimal quantum estimator is also discussed in \cite{Pairs2009}
and \cite{Zhong2014} for a general process. The specific form of the optimal estimator for an su(2) parametrization process will be discussed in our future works.

In summary, we investigate a unitary parametrization process governed
by an su(2) Hamiltonian $\mathbf{r\cdot J}$. We find the optimal input
state and the MQFI when the Hamiltonian is independent of time.
The optimal input state $|\psi\rangle$ in Eq.~\eqref{eq:optimalState} can be a maximal entangled
state for a many-body system. A similar result is discussed in \cite{Fujiwara2001}, where
the author found that the optimal input state for an isotropic depolarization channel
is also a maximally entangled state.
We show that the MQFI can be divided
into two parts. One part is quadratic in  time $t$
due to the dependence of the norm $\mathbf{|r|}$ on the parameter
$\theta$, and the other part oscillates with time $t$ because
of the dependence of the direction $\mathbf{e_r = r/|r|}$ on the
parameter $\theta$.

We apply this result to a typical scenario
and find the MQFI corresponding to the amplitude and the direction
of the magnetic field. We also investigate the MQFI for
an ensemble of atoms interacting with a static field. We find
that in some cases the oscillating terms can contribute
significantly and thus can not be neglected.
We further investigate a time-dependent
driving field and find that the MQFI of the field frequency reaches its optimum
when the driving field is in resonance with the atomic transition
frequency. Moreover, the MQFI of the amplitude of the
external field also reaches its
optimum when the field is in resonance with the atoms.

\begin{acknowledgements}
This work was supported by the NFRPC through Grant
No. 2012CB921602 and by the NSFC through Grant No. 11475146.
H. N. Xiong acknowledges the NSFC through Grants No. 11347220
and No.11404287 and the NSF of Zhejiang Province via Grant No. LQ14A04003.
\end{acknowledgements}

\end{document}